\documentclass[useAMS,usenatbib,usegraphicx]{iau}
\usepackage{graphicx}

\title[IAUS291.~~Gravitational wave memory effect] 
{Search for the gravitational wave memory effect with the Parkes Pulsar Timing Array} 

\author[J. Wang et al.\ ]{
Jingbo Wang$^{1,2,3}$,
G. Hobbs$^{3}$
\and Na Wang$^{1,4}$  
 }
\affiliation{
$^1$Xinjiang Astronomical Observatory, Chinese Academy of Science,  \\
150 Science 1-Street, Urumqi, Xinjiang, China, 830011 
\\ {\tt wangjingbo@xao.ac.cn}  \\[\affilskip]

$^2$University of the Chinese Academy of Sciences, No.19A Yuquan Road, \\Beijing, China, 100049 \\[\affilskip]

$^3$CSIRO Astronomy and Space Science, Australia Telescope National Facility, \\
PO Box 76, Epping, NSW 1710, Australia \\[\affilskip]

$^4$Key Laboratory of Radio Astronomy, Chinese Academy of Science, \\ Nanjing, China, 210008 \\[\affilskip]}
\pubyear{2012}
\volume{291}  
\jname{\mbox{Neutron Stars and Pulsars: Challenges and Opportunities after 80 years}}
\editors{J. van Leeuwen, ed.} 
\begin{document}

\maketitle

\begin{abstract}
Gravitational wave bursts produced by supermassive binary black hole mergers will leave 
a persistent imprint on the space-time metric. Such gravitational wave 
memory signals are detectable by pulsar timing arrays as a glitch event that
would seem to occur simultaneously for all pulsars. In this paper, we describe an initial algorithm which 
can be used to search for gravitational wave memory signals. 
We apply this algorithm to the Parkes Pulsar Timing Array data set. 
No significant gravitational wave memory signal is founded in the data set. 


\end{abstract}
\firstsection 
\section{Introduction}
It is believed that observations of MSPs will 
lead to the direct detection of gravitational waves (GWs) with frequencies of 
$10^{-9}$--$10^{-7}$ Hz (Hobbs et al.\ 2005, Jenet et al.\ 2005). Many observing projects 
have now been started with the goals of observing a 
large enough sample pulsars with sufficient precision to detect GW signals.  Such 
projects are known as pulsar timing arrays (PTAs, Foster 1990). 
Here we make use of data set from the Parkes Pulsar Timing Array project 
(PPTA; Manchester et al.\ 2012). The Parkes observations have already
been used in searching for the GW emission from individual, non-evolving, supermassive black hole binaries 
(Yardley et al.\ 2010), placing an upper limit on a background of GWs (Jenet et al.\ 2006) and  
in attempting to detect such a GW background (Yardley et al.\ 2011). 

In contrast to earlier searches for GWs using the PPTA data sets, in this paper,
we focus on the GW memory (GWM) phenomenon.  The expected source is a supermassive
binary black hole (SMBBH) system that has coalesced (Favata 2009). 
At the coalescence stage of the SMBBH, a permanent change in the space-time metric will
be induced.  Cordes \& Jenet (2012) and van Haasteren \& Levin (2010) have previously shown 
that pulsar timing arrays are sensitive to such GW memory events. When such a GW 
signal passes a pulsar or the Earth, it will lead to a
simple frequency jump in the observed pulse frequency of the pulsar. The timing 
residuals will have the characteristics of a simple glitch event. GW memory
events passing a single pulsar will lead to a glitch-like event in the timing residuals of 
that pulsar only.  GW memory events passing the Earth will lead to a glitch-like event seen
in the timing residuals of all pulsars with the size of the pulse frequency jump depending upon the  
GW source-Earth-pulsar angle. 
The pre-fit timing residuals induced by the GWM signal that occurred at $t=t_0$ can be written as:  
\begin{equation}  
r\left(t\right)_{\rm prefit} = 
\frac{1}{2}h^{\rm mem} (1-\cos{\theta})\cos{2\phi} ~\left(t-t_{0}\right) 
\Theta\left(t-t_{0}\right), 
\end{equation} 
In the above, $\theta$ is the GW source--Earth--pulsar angle, $\phi$ is the angle between the wave's principle 
polarization and the projection of the pulsar onto the plane perpendicular to 
the propagation direction, and h$^{\rm mem}$ is the amplitude of the GWM signal.
Therefore, the pre-fit timing residuals induced by the Earth term of GWM signal will 
give rise to a linear increase of the pre-fit residuals with time. 
\section{Method}

Here we describe our current algorithm and present initial results.  The completed algorithm and our final results will be published elsewhere.  We have updated the \textsc{tempo2} pulsar timing model to include the effect of 
a GWM event. The position of pulsars and the GW source  are specified in the  
equatorial coordinate system by their right ascension and declination ($\alpha$, 
$\delta$). The principle polarisation of GW  
is defined in a coordinate system (r$_g$,$\alpha_g$, 
$\delta_g$) where the GW propagates along the $-r_g$ direction (See
Fig.\ 1 of Hobbs  
et al.\ 2009). Since \textsc{tempo2} only implements a linear least-squares-fitting procedure for 
improving the pulsar timing model, \textsc{tempo2} can only be used to fit for 
the amplitude of the GWM event.  If, as usual, the position, epoch and/or polarisation angle
is unknown and it is necessary to determine these parameters using a different procedure.
A global fitting algorithm (first described in Champion et al.\ 2010) is used to fit 
for  pulse, astrometric and orbital parameters of each pulsars
individually whilst simultaneously fitting for $h^{\rm mem}$.  
In order to account for the unknown polarisation angle (PA) we carry out two 
fits, one with $PA = 0$ and the second with $PA = \pi/4$.  For each fit we 
obtain a measurement of $h^{\rm mem}$ (h$_{1}$ and h$_{2}$ for $PA = 0$ and 
$PA = \pi/4$, respectively) and their 
corresponding uncertainties ($\sigma_{1}$ and $\sigma_{2}$
respectively).  Then we form the following detection statistic: 
\begin{equation} 
S = \left(\frac{h_1}{\sigma_1}\right)^2 + \left(\frac{h_2}{\sigma_2}\right)^2 
\end{equation} 
Coles et al.\ (2011) discussed the issues arising from fitting a pulsar timing model in the 
presence of non-white noise. For this work, we obtain a simple analytic model of the red noise 
for each pulsar (as described in Manchester et al.\ 2012) and use the generalised least-squares-fitting 
routines (often referred to as ``Cholesky" fitting) within \textsc{tempo2}. 
\section{Observations} 
We make use of the  Parkes Pulsar Timing Array (PPTA) data set which is  
described in Manchester et al.\ (2012). These data include regular  
observations of 20 millisecond pulsars at intervals of 2-3 weeks  
from 2005 to 2011. All observations were taken with the  Parkes 64-m radio telescope. 
The typical integration time for each pulsar is about 1 hr.  Most of the timing offsets between 
the different observing systems have been removed.  However, some of the arbitrary jumps from 
the timing model included in the Verbiest et al.\ (2008, 2009) were retained (Manchester et al.\ 2012).  
Variations of dispersion measure were corrected by using multi-frequency observations. 
Timing residuals were formed using the {\sc tempo2} software package (Hobbs et al.\ 2006) 
making use of the JPL DE421 Solar System ephemeris (Folkner et al.\ 2008) and refered to 
terrestrial time as realised by BIPM2011. 

\section{Results} 

We searched for a GWM signal using the algorithm described above. The measured statistic value ranges from 0 to 20 for different trial positions and glitch epochs.  By comparison, the largest detection statistic value is $\sim$ 126 after a small simulated GWM signal is added into the PPTA data set. We therefore conclude that there is not a large GWM signal in the PPTA data set.  We are now carrying out statistical tests of the algorithm in order to answer the following questions: ``what is the largest GWM signal that could be present in our data?" and ``what is the probability that we have already made a detection of a small GWM signal?".   These statistical tests are not yet complete, but are based on Monte-Carlo simulations in which we inject small GWM signals into simulated data sets and measure the detection statistic values. 

It is unlikely that we will make a detection of a GWM event with the existing Parkes data set.  However, the data sets continue to get longer and future analyses are likely to be carried out with a combination of data from different observatories Worldwide.  In the longer term, it is expected that future telescopes such as FAST, Urumqi 110-m and the SKA Phase I will provide data sets in which GWM events will be easily detectable.


~\\
\textbf{Acknowledgments}   \\
This work is supported by National Basic Research Program of China (973 Program 
2009CB824800 and 2012CB821800), NSFC project (No.\ 11173041, No.\ 11173042 and No.\ 11203063), 
the Knowledge Innovation Program 
of the Chinese Academy of Sciences, Grant No. KJCX2-YW-T09, West Light
Foundation of CAS (No.\ XBBS201021 
and No.\ XBBS201123) and Natural Science Foundation of Xinjiang, China Grant No.Y1000201. 
GH acknowledges support from the Australian Research Council \#DP0878388.   

\end{document}